\documentclass[conference]{IEEEtran}

\usepackage{mathrsfs}

\ifCLASSINFOpdf
\else
   \usepackage[dvips]{graphicx}
\fi
\usepackage{url}
\usepackage{cite}
\usepackage{amsmath, amsthm, amssymb, amsfonts}
\usepackage{algorithmic}
\usepackage{graphicx}
\usepackage{textcomp}
\usepackage{relsize}
\usepackage[ruled]{algorithm2e}
\usepackage{algorithmic}
\usepackage{balance}
\usepackage{caption}
\usepackage[labelformat=simple]{subcaption}

\usepackage{cite}
\usepackage{units}
\usepackage{xcolor}
\usepackage{multirow}
\usepackage{booktabs}
\usepackage{arydshln}
\usepackage{epstopdf}
\usepackage{fancyhdr}

\fancypagestyle{plain}{
  \fancyhf{}
  \fancyhead[C]{Conference on \LaTeX}     
  \fancyfoot[L]{This is a notice}

}
\usepackage{eso-pic}

\AddToShipoutPictureBG{%
  \AtPageUpperLeft{%
    \setlength\unitlength{1in}%
    \hspace*{\dimexpr0.5\paperwidth\relax}
    \makebox(0,-0.75)[c]{\textcolor{red}{This article is accepted in the $30^{\text{th}}$ IEEE International Conference on Electronics, Circuits and Systems (ICECS 2023).}}}}

\AddToShipoutPictureBG*{%
  \AtPageLowerLeft{%
    \setlength\unitlength{1in}%
    \hspace{\dimexpr0.07\paperwidth\relax}
    \begin{minipage}[c]{\textwidth}
        {\vspace{-2.2cm}\copyright \small{2023 IEEE. Personal use of this material is permitted.  Permission from IEEE must be obtained for all other uses, in any current or future media,    
    including reprinting/republishing this material for advertising or promotional purposes, creating new collective works, for resale or redistribution to servers or lists, or reuse of any copyrighted component of this work in other works. }}
    \end{minipage}}}
    
\begin{document}
    
\bstctlcite{IEEEexample:BSTcontrol}

\title{A Low Complexity Block-oriented Functional Link Adaptive Filtering Algorithm}


\author{Pavankumar Ganjimala\textsuperscript{1}, Subrahmanyam Mula\textsuperscript{2}\\\normalsize Electrical engineering department, Indian Institute of Technology Palakkad, Palakkad, India\\\normalsize E-mail: \textsuperscript{1}122014005@smail.iitpkd.ac.in, \textsuperscript{2}svmula@iitpkd.ac.in}
\maketitle

\begin{abstract}
The high computation complexity of nonlinear adaptive filtering algorithms poses significant challenges at the hardware implementation level. In order to tackle the computational complexity problem, this paper proposes a novel block-oriented functional link adaptive filter (BO-FLAF) to model memoryless nonlinear systems. Through theoretical complexity analysis, we show that the proposed Hammerstein BO trigonometric FLAF (HBO-TFLAF) has $47\%$ lesser multiplications than the original TFLAF for a filter order of $1024$. Moreover, the HBO-TFLAF exhibits a faster convergence rate and achieved $3-5$ dB lesser steady-state mean square error (MSE) compared to the original TFLAF for a memoryless nonlinear system identification task.

\end{abstract}


\section{Introduction}

Conventionally, signal processing applications such as active noise control (ANC), echo cancellation, channel equalization etc. have relied on linear adaptive filters to model unknown systems due to their ease of design, implementation and lesser computational complexity \cite{nonl_book}. However, in reality, a lot of practical problems involve nonlinear elements \cite{loudspeaker_nonl,gflnn,fiber_nlc} which cannot be modelled by the conventional methods. One way to address this issue is to use nonlinear adaptive filters which incorporate nonlinearity in their model and thus improve the modelling accuracy.

Different structures and algorithms for nonlinear adaptive filters have been proposed in literature. Some of the important classes of algorithms are kernel adaptive filters (KAF) \cite{liu2011kernel}, functional link adaptive filters (FLAF) \cite{flaf} and spline adaptive filters (SAF) \cite{saf}. Out of these the FLAF belongs to a class of filters known as linear-in-the-parameters (LIP) filters and has been widely used in online learning applications \cite{sparse_flaf,hearing_aid}. Most of the recent works improving FLAF \cite{comminiello2015improving,aeflnn,burra2021performance,yu2021robust} focus on improving the performance of the basic FLAF. There are only a few efforts which tackle the high computational complexity issue of nonlinear adaptive filters \cite{anc_tcas1,sicuranza2011bibo,fd-flaf}. Compared to the FLAF, the SAF belongs to the block-oriented class of filters. Block-oriented filters are composed of a cascade of purely linear (L) or nonlinear (NL) modelling blocks and can be of Hammerstein type (NL-L) or Wiener type (L-NL). SAF has been proposed as a low-complexity solution for various applications \cite{saf_anc,saf_tcas1,saf_2d}. The low complexity of SAF can be attributed to its block-oriented structure \cite{bo-tdnn}. With that motivation, we propose a novel low complexity block-oriented FLAF structure in this paper. In the next section, we give a brief overview of the FLAF.


\section{Functional link adaptive filter}
\label{sec:flaf}

\begin{figure}[h]
\centering
\includegraphics[width=8.5cm]{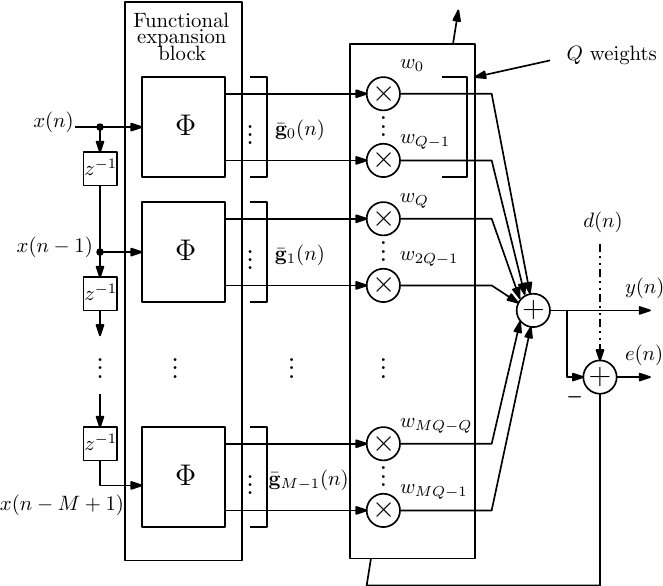}
\caption{Original FLAF}
\label{fig:tflaf_orig}
\end{figure}

The block diagram of an $M$-tap FLAF is shown in Fig. \ref{fig:tflaf_orig}. As depicted in the figure, input samples $x(n)$ are fed to a tapped delay line and each of the $M$ outputs is expanded into $Q$ samples through a functional expansion block (FEB), where $Q$ is the number of functional links. The FEB generates the nonlinear terms which help model the nonlinear system. The functional links are a set of $Q$ functions $\Phi = [\varphi_0(\cdot), \varphi_1(\cdot), \dots \varphi_{Q-1}(\cdot)]$.
The output expansion of an input sample $x(n-i)$ is given as $\bar{\mathbf{g}}_{i}(n) \in \mathbb{R}^Q$, where


\begin{align}
    \bar{\mathbf{g}}_{i}(n) & = \begin{bmatrix}
    \varphi_0(x(n-i)) \\
    \varphi_1(x(n-i)) \\
    \vdots \\
    \varphi_{Q-1}(x(n-i))
    \end{bmatrix}
\end{align}
$i=0, \dots M-1$ and $n$ is the time index. The entire input buffer after expansion results in the expansion buffer $\mathbf{g}(n)$, given as 

\begin{equation}
    \mathbf{g}(n) = [\bar{\mathbf{g}}_{0}(n)^T, \bar{\mathbf{g}}_{1}(n)^T, \dots \bar{\mathbf{g}}_{M-1}(n)^T]^T
\end{equation}

There are various choices for FEB basis functions such as Chebyshev, Legendre and trigonometric polynomials. Among these, the trigonometric polynomial function is widely used in literature as it provides the best compact representation and is computationally inexpensive \cite{flaf}. We consider the trigonometric polynomial in this paper, for which the function expansion of the $i^{\text{th}}$ sample in the tapped delay line is given by 
\begin{equation}
    \varphi_j(x(n-i)) = 
    \begin{cases}
        x(n-i), & j=0\\
        sin(p\pi x(n-i)), & j=2p-1\\
        cos(p\pi x(n-i)), & j=2p
    \end{cases}
\end{equation}
where $p=1\dots P$, and $j=0\dots Q-1$. $P$ is the expansion order which denotes the amount of nonlinearity required and $Q=2P+1$.

If we consider the weight vector $\mathbf{w}(n) = [w_0(n), w_1(n), \dots w_{MQ-1}(n)]$, then the output of the adaptive filter is given as $y(n)=\mathbf{w}(n)^T\mathbf{g}(n)$ and the error $e(n) = d(n)-y(n)$, where $d(n)$ is the desired response. The weight update equation using the stochastic gradient rule is given as
\begin{equation}
    \mathbf{w}(n+1) = \mathbf{w}(n) + \mu e(n)\mathbf{g}(n)
\end{equation}
where $\mu$ is the step size parameter. 

In this section, we described a memoryless FLAF, which models instantaneous nonlinearity. An FLAF with memory can be realized using additional functional expansions \cite{flaf} which can model dynamic nonlinearity, i.e nonlinear functions which depend on the time instant.
\section{Block-Oriented Functional Link Adaptive Filter}
\label{sec:rctflaf}
In this section, we propose two modifications to the FLAF to make it computationally more efficient. First, we describe an improvisation coined as the single $\Phi$ FLAF. Then, we describe the novel block-oriented FLAF algorithm.
\subsection{Single $\Phi$ FLAF}
Instead of delaying the input sample $x(n)$ in a tapped delay line and performing the $M$ $\Phi$ operations in parallel as shown in the original FLAF in Fig. \ref{fig:tflaf_orig}, we can perform a single $\Phi$ operation first on the input sample $x(n)$ and then delay the expanded samples in $\mathbf{g}_{0}(n)$ as shown in Fig. \ref{fig:tflaf_phired}. This idea is already present in nonlinear ANC literature \cite{anc_tcas1,fslms_hw} and we extend this to the FLAF and coin this as the single $\Phi$ architecture.
The $M$ $\Phi$ blocks are identical and time-invariant. Moreover, they just perform a point-wise mapping of the input sample. Hence, this rearrangement does not affect the generation of $\mathbf{g}(n)$. In this topology, $\mathbf{g}(n)=[\Bar{\mathbf{g}}_{0}(n)^T, \Bar{\mathbf{g}}_{0}(n-1)^T, \dots \Bar{\mathbf{g}}_{0}(n-M+1)^T]^T$, where $\Bar{\mathbf{g}}_{0}(n-i)$ is the expanded vector $\Bar{\mathbf{g}}_{0}(n)$ delayed by $i$ samples. In this topology, the number of $\Phi$ modules reduces from $M$ to just $1$ and is independent of filter order $M$, but the number of delay elements increases from $M$ to $MQ$. Generally, delay elements (or memory) are relatively less expensive compared to $\Phi$ block which involves polynomial function evaluations.

%


\subsection{Block-Oriented Functional Link Adaptive Filter}

\begin{figure*}[h]
    \begin{subfigure}[b]{0.35\textwidth}
        \centering
        \includegraphics[width=7.5cm]{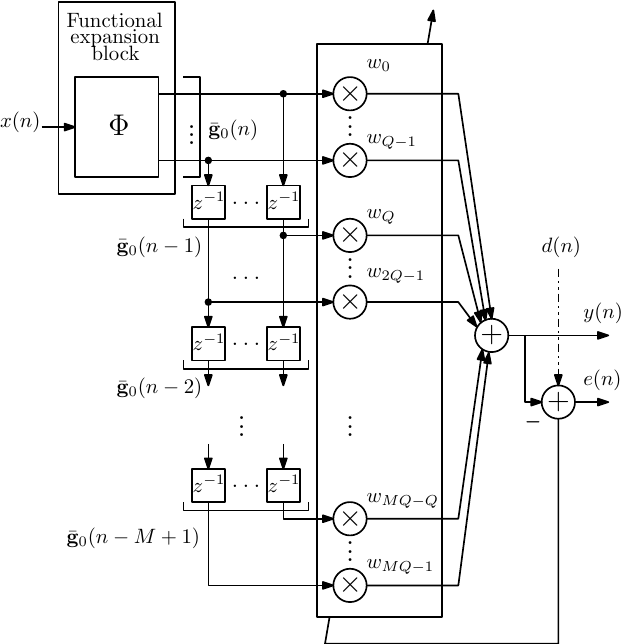}
        \caption{FLAF with single $\Phi$ module}
        \label{fig:tflaf_phired}
    \end{subfigure}
    \hfill
    \begin{subfigure}[b]{0.55\textwidth}
        \centering
        \includegraphics[width=9.5cm]{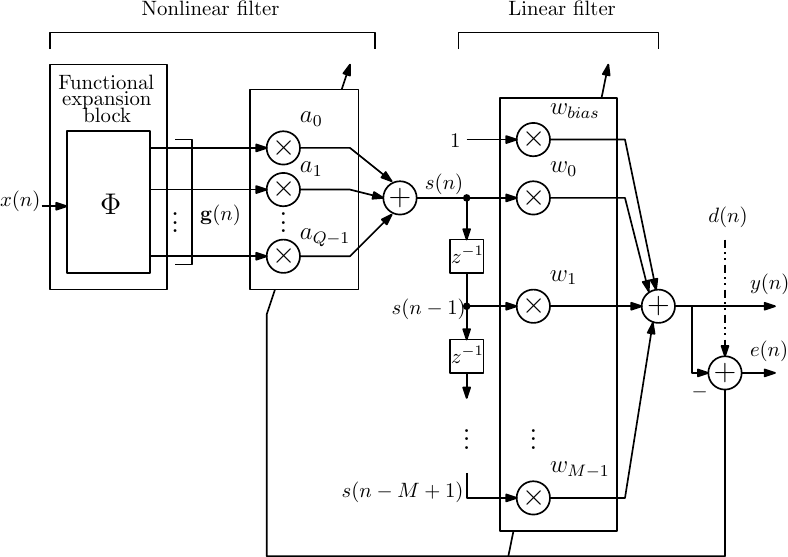}
        \caption{Hammerstein block-oriented FLAF (HBO-FLAF) structure}
        \label{fig:rctflaf_diag}
    \end{subfigure}
    \caption{Proposed FLAF modifications}
\end{figure*}

Here we further reduce the computational complexity of the single $\Phi$ FLAF by designing a block-oriented Hammerstein structure inspired from the SAF structure \cite{hammerstein_saf}. We split the parallel $MQ$ tap filter in FLAF into two separate serial filters of order $M$ and $Q$ respectively and create a novel Hammerstein type filter called the Hammerstein block-oriented FLAF (HBO-FLAF) which is shown in Fig. \ref{fig:rctflaf_diag}. In the first stage the $Q$ samples from the FEB are adaptively combined by the adaptive weights $\mathbf{a}(n) = [a_0(n), a_1(n), \dots a_{Q-1}(n)]^T$ to identify the nonlinear component of the system. The output of the first stage $s(n)$ is given to a tapped delay line of length $M$ and the outputs are then adaptively combined by the second filter with weights $\mathbf{w}(n) = [w_{bias}(n),w_0(n), w_1(n), \dots w_{M-1}(n)]^T$ to identify the linear component. Here $w_{bias}$ is an optional adaptive bias quantity added to obtain improved filter performance.

The output of the nonlinear filter $s(n)$ is given by 
\begin{equation}
    s(n) = \mathbf{a}(n)^T\mathbf{g}(n)
    \label{eq:nonlin_op}
\end{equation}
and $\mathbf{g}(n) = \Bar{\mathbf{g}}_0(n) = [\varphi_0(x(n)), \varphi_1(x(n)), \dots \varphi_{Q-1}(x(n))]^T$

The final output $y(n)$ is given by

\begin{equation}
    y(n) = \mathbf{w}(n)^T\mathbf{s}(n)
    \label{eq:lin_op}
\end{equation}
where, $\mathbf{s}(n) = [1, s(n), s(n-1), \dots s(n-M+1)]^T$ is a buffer consisting of the first stage outputs. The estimation error is defined as $e(n) = d(n) - y(n)$, and $d(n)$ is the desired signal.





The HBO-FLAF has two weight updates to be performed. The weight update equation is derived here using the standard stochastic gradient descent method \cite{sayed2003fundamentals}. The weight update for the linear filter weights $\mathbf{w}(n+1)$ using the MSE cost function can be derived as follows

\begin{equation*}
    \begin{split}
        \mathbf{w}(n+1) & = \mathbf{w}(n)+ \frac{\partial(e(n)^2)}{\partial \mathbf{w}(n)}\\
                        & = \mathbf{w}(n)+ 2e(n)\frac{\partial(d(n)-\mathbf{w}(n)^T\mathbf{s}(n))}{\partial \mathbf{w}(n)}\\
    \end{split}
    \label{eq:gen_wupd_lms}
\end{equation*}

Simplifying the derivative and replacing the constant terms with a learning rate parameter $\mu_w$ \cite{saf}, we get the final weight update equation for the linear weights $\mathbf{w}(n)$ as 

\begin{equation}
    \mathbf{w}(n+1) = \mathbf{w}(n) + \mu_w e(n)\mathbf{s}(n)
    \label{eq:w_upd_th}
\end{equation}
Similarly, the weight update for the nonlinear filter weights $\mathbf{a}(n+1)$ can be derived using stochastic gradient descent as 

\begin{equation}
        \mathbf{a}(n+1) = \mathbf{a}(n)+ 2e(n)\frac{\partial(d(n)-\mathbf{w}(n)^T\mathbf{s}(n))}{\partial \mathbf{a}(n)}
\label{eq:a_deriv1}
\end{equation}
Rewriting $\mathbf{s}(n)$ in terms of $\mathbf{a}(n)$

\begin{equation*}
    \mathbf{s}(n)^T = [\mathbf{a}(n)^T\mathbf{g}(n), \mathbf{a}(n-1)^T\mathbf{g}(n-1), \dots \mathbf{a}(n-M+1)^T\mathbf{g}(n-M+1)]
\end{equation*}

For a small step size, we can assume that the weights $\mathbf{a}(n)$ change very little over each iteration \cite{sayed2003fundamentals}, i.e $\mathbf{a}(n)\approx\mathbf{a}(n-1)\approx\dots\mathbf{a}(n-M+1)$. Then $\mathbf{s}(n)$ can be written as 
\begin{equation}
\begin{aligned}[b]
    \mathbf{s}(n)^T & = [\mathbf{a}(n)^T\mathbf{g}(n), \dots \mathbf{a}(n)^T\mathbf{g}(n-M+1)]\\
    & = \mathbf{a}(n)^T[\mathbf{g}(n), \mathbf{g}(n-1), \dots \mathbf{g}(n-M+1)]\\
    & = \mathbf{a}(n)^T\mathbf{G}^T
    \label{eq:sn_int}
\end{aligned}
\end{equation}
where $\mathbf{G}^T$ is a $Q\times M$ matrix defined as 
\begin{equation}
    \mathbf{G}^T = [\mathbf{g}(n), \mathbf{g}(n-1), \dots \mathbf{g}(n-M+1)]
\end{equation}

Substituting (\ref{eq:sn_int}) in (\ref{eq:a_deriv1}) we get 
\begin{equation*}
       \mathbf{a}(n+1) = \mathbf{a}(n)+ 2e(n)\frac{\partial(d(n)-\mathbf{w}(n)^T\mathbf{G}\cdot\mathbf{a}(n))}{\partial \mathbf{a}(n)}
\end{equation*}

Similar to (\ref{eq:w_upd_th}), we simplify the equation for the nonlinear weights $\mathbf{a}(n)$ using learning rate parameter $\mu_a$ as

\begin{equation}
    \mathbf{a}(n+1) = \mathbf{a}(n) + \mu_a e(n)\mathbf{w}(n)^T\mathbf{G}
    \label{eq:a_upd_th}
\end{equation}

In this section, we presented a memoryless version of the HBO-TFLAF. Similar to FLAF with memory, HBO-FLAF is capable of modelling nonlinear systems with memory by adding a delay line to $\mathbf{g}(n)$. Additional adaptive weights are added to the delay line samples and their cross terms.

\subsection{Theoretical computational complexity}
The computational complexity of an algorithm can be measured in terms of the number of arithmetic operations performed in an iteration. The computational complexity of the least mean squares (LMS), trigonometric FLAF (TFLAF), single $\Phi$ TFLAF, HBO-TFLAF and Hammerstein SAF (HSAF) algorithms are shown in Table \ref{tab:complexity}. In this study, we only consider the memoryless version of the TFLAF algorithm. Here we define $M$ as the length of the linear filter, $Q_t$ and $P_t$ as the $Q$ and $P$ parameters in TFLAF, HBO-TFLAF and $Q_h=P_h+1$, where $P_h$ is the order of nonlinear function in HSAF. 

The TFLAF algorithm has $Q_t$ times more multiplications and additions compared to the LMS algorithm. In addition, TFLAF also has $M(Q_t-1)$ trigonometric function evaluations and $MP_t$ multiplications in the FEB. The single $\Phi$ structure reduces the number of trigonometric function evaluations to $Q_t-1$ and correspondingly the $MP_t$ multiplications in the FEB to $P_t$ multiplications. The HBO-TFLAF further reduces the computation by splitting the $MQ_t$ tap filter into two filters of order $M$ and $Q_t$ respectively ($M+Q_t$ feed-forward taps). Although the HBO-TFLAF has only $M+Q_t$ forward taps, the additional vector-matrix multiplication between $\mathbf{w}(n)$ and $\mathbf{G}$ in (\ref{eq:a_upd_th}) contributes to $MQ_t$ multiplications and $(M-1)Q_t$ additions. The computational complexity of HSAF is similar to the HBO-TFLAF and has $Q_h$ instead of $Q_t$ taps.

\begin{table}[h]
    \centering
    \caption{Theoretical computational complexity}
    \label{tab:complexity}    
    \begin{tabular}{|c|c|c|c|c|}
    \hline
        Algorithm & Number of & Number of & Trig.\\
                  & multipliers & adders & functions  \\ 
        \hline
        \hline
        LMS & $2M + 1$ & $2M$ & 0\\
        \hline
        TFLAF & $2MQ_t + MP_t + 1$ & $2MQ_t$ & $M(Q_t-1)$\\
        \hline
        Single $\Phi$ TFLAF & $2MQ_t + P_t + 1$ & $2MQ_t$ & $Q_t-1$\\
        \hline
        \multirow{2}{*}{HBO-TFLAF} & $2(M+Q_t)+ P_t$ & $2M+Q_t+$ & \multirow{2}{*}{$Q_t-1$}\\
         & $+MQ_t+1$ & $MQ_t-1$ & \\
        \hline
        \multirow{3}{*}{HSAF} & $2(M+Q_h)+P_h$ & $2M+Q_h+$ & \multirow{3}{*}{$0$}\\
         & $+MQ_h$ & $MP_h+Q_h+$ & \\
         & $+(P_h+1)^2$ & $3+P_h(P_h+1)$ & \\
        \hline
    \end{tabular}

\end{table}

\begin{figure*}[h!]
     \centering
      \begin{subfigure}[b]{0.32\textwidth}
         \centering
         \includegraphics[width=\textwidth]{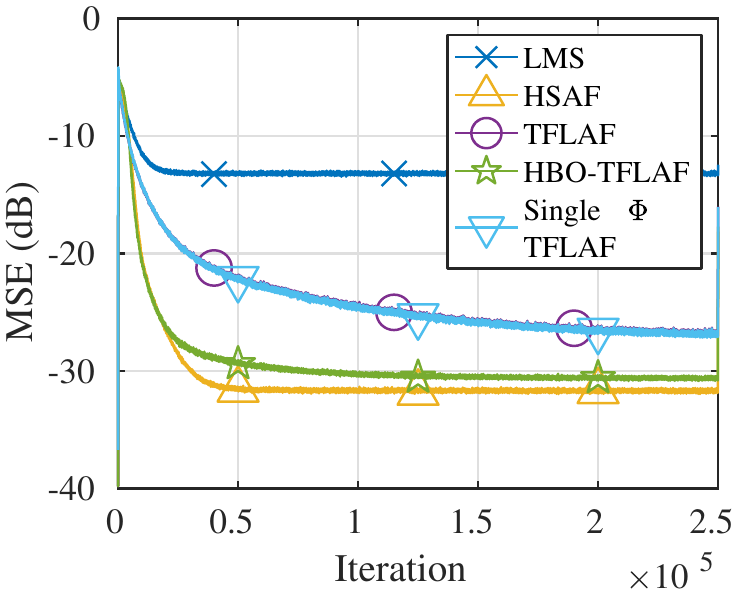}
         \caption{Memoryless system 1}
         \label{fig:tran_chara}
     \end{subfigure}
     \hfill
      \begin{subfigure}[b]{0.32\textwidth}
         \centering
         \includegraphics[width=\textwidth]{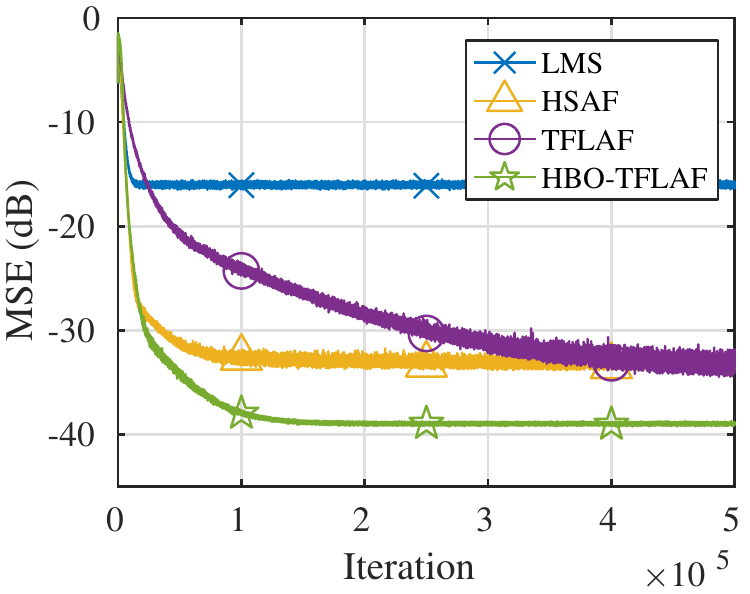}
         \caption{Memoryless system 2}
         \label{fig:soft_clip_chara}
     \end{subfigure}
     \hfill
      \begin{subfigure}[b]{0.32\textwidth}
         \centering
          \includegraphics[width=\textwidth]{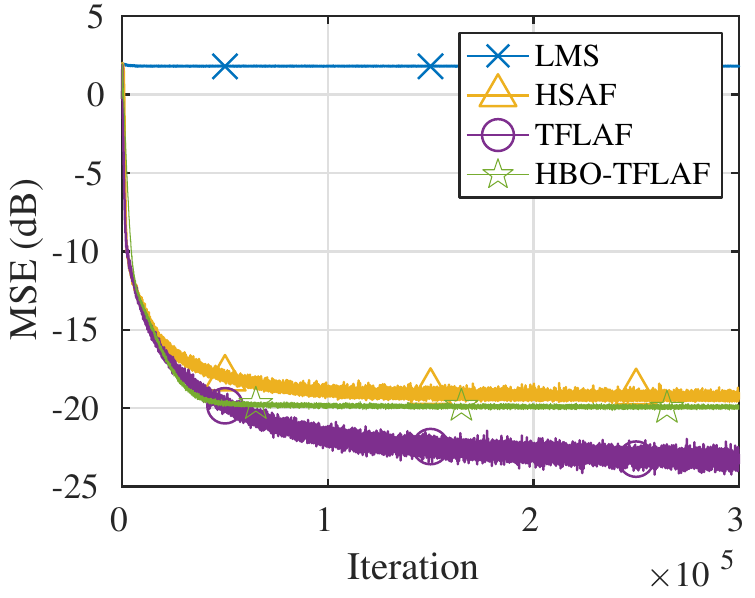}
         \caption{System with memory}
         \label{fig:memoryless_time_diff}
     \end{subfigure}
        \caption{MSE learning curves}
        \label{fig:mse_del}
\end{figure*}

To obtain a rough estimate of the computational complexity, we consider an acoustic echo cancellation application and assume $M=1024$ and $Q_t=7$, $Q_h=4$. Substituting the values, the number of multiplications for TFLAF, single $\Phi$ TFLAF, HBO-TFLAF and HSAF are $17409,14340,9235$ and $6171$ respectively and the number of additions are $14336, 14336, 9222$ and $5143$ respectively. The number of multiplications and additions is in the order: TFLAF $>$ Single $\Phi$ TFLAF $>$ HBO-TFLAF $>$ HSAF. Although the HBO-TFLAF is similar in structure to HSAF, since $Q_t>Q_h$, the HBO-TFLAF has higher computation. HBO-TFLAF will have lower computation than the HSAF when $Q_t<Q_h$. 

\section{MSE Performance Analysis}
\label{sec:mat_res}


In order to assess the performance of the TFLAF, single $\Phi$ TFLAF and HBO-TFLAF, we perform nonlinear system identification on three different systems in MATLAB. We also compare performance with the HSAF and the LMS algorithm which is a linear model. The performance metric chosen here is the mean square error (MSE) metric given by $10 \cdot\text{log}_{10}(E[e(n)^2])$. The MSE learning curves are generated from averaging $500$ independent Monte Carlo experiments followed by a $20$ tap moving average filter to improve the visibility of the curves.



Input to the systems is white Gaussian with a variance of $0.25$ and noise variance is taken to be $0.01$. The filter weights in LMS, RFF-KLMS, TFLAF and linear filter weights in HSAF and HBO-TFLAF are initialized to zeros. The nonlinear adaptive weight vector $\mathbf{a}(n)$ in HBO-TFLAF is initialized as $\delta(n)$, where the first weight, which corresponds to the $x(n)$ output is one and the rest are zeros. This ensures that the HBO-TFLAF structure behaves as a linear filter initially. The bias weight ($w_{bias}$) is also used for the TFLAF and HBO-TFLAF. The HSAF nonlinear weight vector denoted by $\mathbf{q}(n)$ is set as $[-3.0, -2.75,\dots 2.75, 3]$ (for $\Delta x = 0.25$) resembling a linear mapping. In HSAF, for all the experiments $Q_h=4$ and the matrix $C$ is chosen as the CR-spline basis matrix (HSAF parameter definitions in \cite{hammerstein_saf}). The parameters chosen for the various filters in the three systems are shown in Table \ref{tab:parameters}. The inputs and the weight initialization are kept the same for all the experiments.

\begin{table}[h]
    \caption{Experiment parameters}
    \label{tab:parameters}
    \centering
    \begin{tabular}{|c|c|c|c|}
    \hline
         \multirow{2}{*}{Parameter} & Memoryless & Memoryless & System\\
         & system $1$ & system $2$ & with memory \\
         \hline
         \hline
         $M$ & $512$ & $512$ & $8$\\
         \hline
         $\mu_{LMS}$ & $0.002$ & $0.004$ & $0.003$\\ 
         \hline
         $\mu_{TFLAF}$ & $0.0005$ & $0.0004$ & $0.0002$\\ 
         \hline
         $\mu_{w-BOTFLAF}$ & $0.0006$  & $0.0006$ & $0.0004$\\ 
         \hline
         $\mu_{a-BOTFLAF}$ & $0.0006$  & $0.0011$ & $0.0005$\\ 
         \hline
         $Q_{t}$ & $9$ & $9$ & $7$\\ 
         \hline
         $\mu_{w-HSAF}$ & $0.0015$  & $0.0018$ & $0.002$\\ 
         \hline
         $\mu_{q-HSAF}$ & $0.0015$ & $0.0075$ & $0.005$\\ 
         \hline
         $\Delta x$ & $0.25$ & $0.21$ & $0.25$\\
         \hline
    \end{tabular}
\end{table}









\subsubsection{Memoryless systems}

First, we consider the system identification of Hammerstein-type memoryless nonlinear systems. In system $1$, an asymmetric loudspeaker distortion system \cite{flaf} is considered. The nonlinear system is cascaded with a linear system which is a reverberation effect generated using the image source method (ISM) with a reverberation time of $T_{60} = 60$ ms and a sampling rate of $8$ kHz and is truncated after $512$ samples.

The MSE learning curves for system $1$ are shown in Fig. \ref{fig:tran_chara}. The nonlinear adaptive filters perform better compared to the LMS algorithm. The HBO-TFLAF, in spite of having lesser complexity than the TFLAF, performs better than TFLAF/Single $\Phi$ TFLAF and obtains around $3$ dB lower MSE. The performance of HBO-TFLAF and HSAF is almost similar, with HSAF obtaining around $1$ dB lower MSE here. We can also see that the learning curves of single $\Phi$ TFLAF and the TFLAF overlap and this proves that the single $\Phi$ TFLAF structure is functionally identical to the TFLAF. In the subsequent plots, we omit single $\Phi$ TFLAF learning curves.

Next, we perform the system identification of another memoryless system, system $2$ which is a soft-clipping distortion system as described in \cite{sparse_flaf}, where we consider $\zeta = 0.35$. The linear system is kept the same as before. The MSE learning curves for system $2$ are shown in Fig. \ref{fig:soft_clip_chara}. The performance is similar to system $1$, and the HBO-TFLAF outperforms all the algorithms and obtains $5$ dB lower MSE compared to the HSAF in this case.

From the results of the two memoryless systems, we conclude that the block-oriented algorithms HBO-TFLAF/HSAF are the best choice to model memoryless systems. Since TFLAF has more degrees of freedom, it tends to over-fit and thus has lesser performance than HBO-TFLAF/HSAF. The lesser number of taps in HBO-TFLAF/HSAF also allows it to converge to the solution faster. Although HSAF and HBO-TFLAF are similar in structure, we observe that HBO-TFLAF has a global weight adaptation behaviour as opposed to the local adaptation of HSAF. The HSAF would therefore need samples spread across the input range to learn the entire nonlinearity, whereas the HBO-TFLAF would not need such a broad range of training samples to learn the entire nonlinearity. From the experiments, we observe that the HBO-TFLAF models symmetric and regularly shaped nonlinearities better, whereas HSAF is better at modelling irregular nonlinearities. A detailed study comparing the HSAF and HBO-TFLAF is beyond the scope of this paper and is necessary to reach more solid conclusions. Additionally, we note that the HBO-TFLAF is slightly easier to set up and function as it has only $4$ hyperparameters as opposed to $5$ in HSAF.

\subsubsection{System with memory}

Next, to understand the limitation of block-oriented systems, we test their MSE performance in identifying a nonlinear system with memory. We consider the following system whose output $y(n)$, given an input $x(n)$ is
\begin{multline}
    y(n) = 0.6sin(\pi x(n))^3 + 0.2cos(2\pi x(n-2))^2 \\
    - 0.1cos(4\pi x(n-4)) + 1.125
\end{multline}

This type of system depends nonlinearly on an input sample and its past samples (or has memory) but does not contain cross terms between the current and past samples. We point out here that the definition of a system with memory is slightly different in \cite{flaf}, which defines it as systems which have cross terms between samples in memory. 

The MSE learning curves for this system are shown in Fig. \ref{fig:memoryless_time_diff}. The TFLAF has the lowest MSE for this system. TFLAF can adapt the weights corresponding to the functional expansion of each input sample in the memory buffer separately. Whereas, the block-oriented models combine the functional expansion before the memory buffer and therefore will not be able to model systems with memory. As mentioned earlier, it is possible to add a delay line and additional filter weights to improve HBO-TFLAF performance for memory systems. With added filter weights, the computational advantage of HBO-TFLAF comes down. The single $\Phi$ TFLAF becomes more computationally efficient for larger order of memory.

\section{Conclusion and Future work}

In this paper, a novel nonlinear adaptive filter structure coined HBO-FLAF is proposed. HBO-FLAF breaks down the original FLAF into two stages and performs the nonlinear and linear modelling separately. Through theoretical analysis and MATLAB simulations, we demonstrated that HBO-TFLAF not only has a lower computational complexity compared to TFLAF but it also achieves faster convergence and lesser steady-state MSE compared to TFLAF for a memoryless nonlinear system identification task. Therefore, HBO-FLAF can be a potential candidate for real-time VLSI applications involving nonlinear system identification. We also introduced the single $\Phi$ FLAF structure which has lower computational complexity than TFLAF and maintains identical performance as TFLAF. In the future, we plan to implement the proposed algorithms in hardware and study the power, performance and area trade-offs. 


\section*{Acknowledgement}
Pavankumar Ganjimala's work was supported by the Prime Minister’s Research Fellowship (PMRF), Ministry of Education (MoE), Government of India.

\bibliographystyle{IEEEtran}\bibliography{main_bib_short}

\begin{thebibliography}{10}
\providecommand{\url}[1]{#1}
\csname url@samestyle\endcsname
\providecommand{\newblock}{\relax}
\providecommand{\bibinfo}[2]{#2}
\providecommand{\BIBentrySTDinterwordspacing}{\spaceskip=0pt\relax}
\providecommand{\BIBentryALTinterwordstretchfactor}{4}
\providecommand{\BIBentryALTinterwordspacing}{\spaceskip=\fontdimen2\font plus
\BIBentryALTinterwordstretchfactor\fontdimen3\font minus \fontdimen4\font\relax}
\providecommand{\BIBforeignlanguage}[2]{{%
\expandafter\ifx\csname l@#1\endcsname\relax
\typeout{** WARNING: IEEEtran.bst: No hyphenation pattern has been}%
\typeout{** loaded for the language `#1'. Using the pattern for}%
\typeout{** the default language instead.}%
\else
\language=\csname l@#1\endcsname
\fi
#2}}
\providecommand{\BIBdecl}{\relax}
\BIBdecl

\bibitem{nonl_book}
D.~Comminiello and J.~C. Principe, \emph{Adaptive Learning Methods for Nonlinear System Modeling}.\hskip 1em plus 0.5em minus 0.4em\relax Butterworth-Heinemann, 2018.

\bibitem{loudspeaker_nonl}
W.~Klippel, ``tutorial: loudspeaker nonlinearities—causes, parameters, symptoms,'' \emph{journal of the audio engineering society}, vol.~54, no.~10, pp. 907--939, october 2006.

\bibitem{gflnn}
G.~L. Sicuranza and A.~Carini, ``{A Generalized FLANN Filter for Nonlinear Active Noise Control},'' \emph{IEEE Trans. on Audio, Speech, and Language Process.}, vol.~19, no.~8, pp. 2412--2417, 2011.

\bibitem{fiber_nlc}
A.~Amari, O.~A. Dobre, R.~Venkatesan \emph{et~al.}, ``A survey on fiber nonlinearity compensation for 400 gb/s and beyond optical communication syst.'' \emph{IEEE Communications Surveys \& Tutorials}, vol.~19, no.~4, pp. 3097--3113, 2017.

\bibitem{liu2011kernel}
W.~Liu, J.~C. Principe, and S.~Haykin, \emph{Kernel adaptive filtering: a comprehensive introduction}.\hskip 1em plus 0.5em minus 0.4em\relax John Wiley \& Sons, 2011, vol.~57.

\bibitem{flaf}
D.~Comminiello, M.~Scarpiniti, L.~A. Azpicueta-Ruiz \emph{et~al.}, ``Functional link adaptive filters for nonlinear acoustic echo cancellation,'' \emph{IEEE Trans. on Audio, Speech, and Language Process.}, vol.~21, no.~7, pp. 1502--1512, 2013.

\bibitem{saf}
M.~Scarpiniti, D.~Comminiello, R.~Parisi \emph{et~al.}, ``Nonlinear spline adaptive filtering,'' \emph{Signal Process.}, vol.~93, no.~4, pp. 772--783, 2013.

\bibitem{sparse_flaf}
D.~Comminiello, M.~Scarpiniti, L.~A. Azpicueta-Ruiz \emph{et~al.}, ``Nonlinear acoustic echo cancellation based on sparse functional link representations,'' \emph{IEEE/ACM Trans. on audio, speech, and language Process.}, vol.~22, no.~7, pp. 1172--1183, 2014.

\bibitem{hearing_aid}
Vasundhara, N.~B. Puhan, and G.~Panda, ``De-correlated improved adaptive exponential flaf-based nonlinear adaptive feedback cancellation for hearing aids,'' \emph{IEEE Trans. Circuits Syst. I: Reg. Papers}, vol.~65, no.~2, pp. 650--662, 2018.

\bibitem{comminiello2015improving}
D.~Comminiello, M.~Scarpiniti, S.~Scardapane \emph{et~al.}, ``Improving nonlinear modeling capabilities of functional link adaptive filters,'' \emph{Neural Networks}, vol.~69, pp. 51--59, 2015.

\bibitem{aeflnn}
V.~Patel, V.~Gandhi, S.~Heda \emph{et~al.}, ``Design of adaptive exponential functional link network-based nonlinear filters,'' \emph{IEEE Trans. Circuits Syst. I: Reg. Papers}, vol.~63, no.~9, pp. 1434--1442, 2016.

\bibitem{burra2021performance}
S.~Burra and A.~Kar, ``Performance analysis of an improved split functional link adaptive filtering algorithm for nonlinear aec,'' \emph{Applied Acoustics}, vol. 176, p. 107863, 2021.

\bibitem{yu2021robust}
T.~Yu, W.~Li, Y.~Yu \emph{et~al.}, ``Robust adaptive filtering based on exponential functional link network: Analysis and application,'' \emph{IEEE Trans. Circuits Syst. II: Express Briefs}, vol.~68, no.~7, pp. 2720--2724, 2021.

\bibitem{anc_tcas1}
D.~Zhou and V.~DeBrunner, ``Efficient adaptive nonlinear filters for nonlinear active noise control,'' \emph{IEEE Trans. Circuits Syst. I: Reg. Papers}, vol.~54, no.~3, pp. 669--681, 2007.

\bibitem{sicuranza2011bibo}
G.~L. Sicuranza and A.~Carini, ``On the bibo stability condition of adaptive recursive flann filters with application to nonlinear active noise control,'' \emph{IEEE Trans. on Audio, Speech, and Language Process.}, vol.~20, no.~1, pp. 234--245, 2011.

\bibitem{fd-flaf}
D.~Comminiello, A.~Nezamdoust, S.~Scardapane \emph{et~al.}, ``A new class of efficient adaptive filters for online nonlinear modeling,'' \emph{IEEE Trans. on Syst., Man, and Cybernetics: Syst.}, 2022.

\bibitem{saf_anc}
V.~Patel and N.~V. George, ``Nonlinear active noise control using spline adaptive filters,'' \emph{Applied Acoustics}, vol.~93, pp. 38--43, 2015.

\bibitem{saf_tcas1}
P.~P. Campo, A.~Brihuega, L.~Anttila \emph{et~al.}, ``Gradient-adaptive spline-interpolated lut methods for low-complexity digital predistortion,'' \emph{IEEE Trans. Circuits Syst. I: Reg. Papers}, vol.~68, no.~1, pp. 336--349, 2020.

\bibitem{saf_2d}
C.~Liu and H.~Zhao, ``A 2d-lut scheme design for complex-valued spline adaptive filter,'' \emph{IEEE Trans. Circuits Syst. II: Express Briefs}, pp. 1--1, 2023.

\bibitem{bo-tdnn}
C.~Jiang, H.~Li, W.~Qiao \emph{et~al.}, ``Block-oriented time-delay neural network behavioral model for digital predistortion of rf power amplifiers,'' \emph{IEEE Trans. Microw. Theory Tech.}, vol.~70, no.~3, pp. 1461--1473, 2022.

\bibitem{fslms_hw}
B.~K. Mohanty, G.~Singh, and G.~Panda, ``Hardware design for vlsi implementation of fxlms-and fslms-based active noise controllers,'' \emph{Circuits, Syst., and Signal Process.}, vol.~36, no.~2, pp. 447--473, 2017.

\bibitem{hammerstein_saf}
M.~Scarpiniti, D.~Comminiello, R.~Parisi \emph{et~al.}, ``Hammerstein uniform cubic spline adaptive filters: Learning and convergence properties,'' \emph{Signal Process.}, vol. 100, pp. 112--123, 2014.

\bibitem{sayed2003fundamentals}
A.~H. Sayed, \emph{Fundamentals of adaptive filtering}.\hskip 1em plus 0.5em minus 0.4em\relax John Wiley \& Sons, 2003.

\end{thebibliography}

\end{document}